# Real-time Peer-to-Peer Botnet Detection Framework based on Bayesian Regularized Neural Network

Sharath Chandra Guntuku, Pratik Narang*, Chittaranjan Hota

Information Security Group,

Department of Computer Science,

Birla Institute of Technology and Science, Pilani,

Hyderabad Campus, Hyderabad, Andhra Pradesh, India

{f2009149, p2011414, hota}@hyderabad.bits-pilani.ac.in

*Corresponding author

## ABSTRACT

Over the past decade, the Cyberspace has seen an increasing number of attacks coming from botnets using the Peer-to-Peer (P2P) architecture. Peer-to-Peer botnets use a decentralized Command & Control architecture. Moreover, a large number of such botnets already exist, and newer versions- which significantly differ from their parent bot- are also discovered practically every year. In this work, the authors propose and implement a novel hybrid framework for detecting P2P botnets in live network traffic by integrating Neural Networks with Bayesian Regularization. Bayesian Regularization helps in achieving better generalization of the dataset, thereby enabling the detection of botnet activity even of those bots which were never used in training the Neural Network. Hence such a framework is suitable for detection of newer and unseen botnets in live traffic of a network. This was verified by testing the Framework on test data unseen to the Detection module (using untrained botnet dataset), and the authors were successful in detecting this activity with an accuracy of 99.2 %.

## 1. PEER-TO-PEER BOTNETS

A Bot is a program written to covertly access the machine on which it is installed, to allow an unauthorized user to remotely control the targeted system (victim) through communication protocols (IRC, HTTP or P2P). A network of such compromised end-hosts under the remote command of a bot-master (or a bot-herder) is called a botnet. The types of attacks that the network gets exposed to are enormous, considering the capability of the bots to autonomously and automatically run on the host computers. Table 1 lists the attacks that are possible to deploy using Botnets.

Table 1. Attacks possible with botnets

| Attack | Description |
|---|---|
| Distributed Denial of Service | Overloading and preventing a single system (hosting a critical application) from servicing legitimate requests |
| Adware | Active advertising of a commercial offering without the user's permission or awareness |
| Spyware | Sending information to the botmasters about a victim's activities – typically credit card numbers, passwords and other information that can be sold on the black market. |
| Email Spam | Flooding people with mails disguised as messages from people but containing advertisements or malicious links |
| Fast-flux | A DNS technique used by botnets to hide malware delivery and phishing sites behind an ever-changing network of compromised hosts acting as proxies |
| Scareware | Inducing users to buy a rogue anti-virus to regain access of their corrupted system |

Botnets using IRC as the channel for Command and Control (C & C) are the ones which can be detected most easily. This is because of the characteristics of the communication channel that IRC provides. Only one central C & C server exists in an IRC based Botnet, and if the server is incapacitated, the entire Botnet is taken down. This single point of failure in IRC botnets led to the emergence of 'smarter' Botnets, such as those which use P2P architecture for Botnet communication. P2P architecture for botnets, shown in figure 1b, is tolerant to the single point of failure as any node in the P2P network can act as both a client and a server. Even if one or two malicious nodes in the P2P Botnet are taken down, the gaps in the network overlay are filled by readjusting the architecture and the network continues to operate under the control of the attacker, shown in figure 1b.

Although P2P bots were on the rise since early 21st century, with the variants of Agobot, Spybot and Sinit launching many exploits on the internet, the task of detecting and mitigating them remains a challenge. Table 2 shows a timeline of P2P Bots, which are continuing to evolve till date introducing complex evolutionary mechanisms to make themselves robust:

Table 2. A timeline of P2P Botnet

| Year of detection | Name of P2P Bot | Technology used |
|---|---|---|
| 2002 | Agobot, Spybot | Early botnets exploiting P2P architecture |
| 2003 | Sinit | Finding peers through random scanning |
| 2004 | Phatbot | Based on WASTE – small VPN-style network with RSA public keys |
| 2006 | Spamthru; Nugache | Advances in custom backup protocol and peer location |
| 2007 | Peacomm; Storm | Exploiting existing Kademlia network |
| 2008 | Conficker; Sality | Dictionary attacks on administrative passwords; Incorporation of evolutionary rootkit functions |
| 2009 | TDL4; Zeus | Encrypted communication and infecting MBR of victims; Man in the browser keystroke logging and Form Garbing |
| 2010 | Waledac; Kelihos; Stuxnet | Email Spam; Theft of bitcoins; Programmable Logic Controller rootkit |
| 2013 | Wordpress | Brute-force crack administrative credentials |

Even though the P2P architecture introduces latency into the communication network, it enables the Botnet to remain hidden. When a bot receives a message, it randomly searches for the next bot in the P2P network and passes the message onto the peer, thereby giving away very less information about the Botnet infrastructure to the outside world.

Recently social networks have also been used as a medium for commanding the bots by the bot-masters. Users of a social networks will generally trust the links and messages coming from the network. This fact was exploited by the Svelta malware *(Wood 2010)* where the Botnet used a twitter account named upd3t3, and commands were sent as tweets from this account in base64 encoded form. The infected bots were the followers of this twitter account, which were programmed to convert the base64 encoded tweets into commands to be executed.

There have also been anonymous groups which asked for volunteers to opt-in to becoming a part of a politically-based Botnet activity. One group in 2010, comprising of 30,000 bots, announced its support for Wikileaks by launching DDoS attacks against anti-Wikileaks companies. *(BBC News 2012)*

Botnet activity was also found in mobile devices, smartphones and tablet PCs in the recent times. Zeus was observed on many Nokia phones which use Symbian OS, which defeated the online banking two-step authentication by monitoring the SMS sent by the bank. *(Apvrille 2012)*

Thus P2P Botnet detection and mitigation is proving to be an extremely important area of research both on the desktop and the mobile platforms. Research on how they recruit bot members, form the botnet and finally attack is given in *(Wang, et al. 2009)*.

In this paper the authors propose a novel hybrid framework by integrating Neural Networks with a Bayesian Regularization pre-processing module. Bayesian Regularization helps in achieving better generalization of the dataset, thereby enabling the detection of botnet activity of even those bots which were never used in training the Neural Network. This means that such a framework is suitable for detection of newer and unseen botnets in live traffic of a network, as evident by the results of this research. Owing to the generalization provided by Bayesian Regularization, the authors were successful in detecting activity of untrained malicious bots with an accuracy of 99.2 %. This model was then integrated into a Java API, which can be used as a pre-processor module for any Intrusion Detection System for the real-time detection of any botnet traffic on the network.

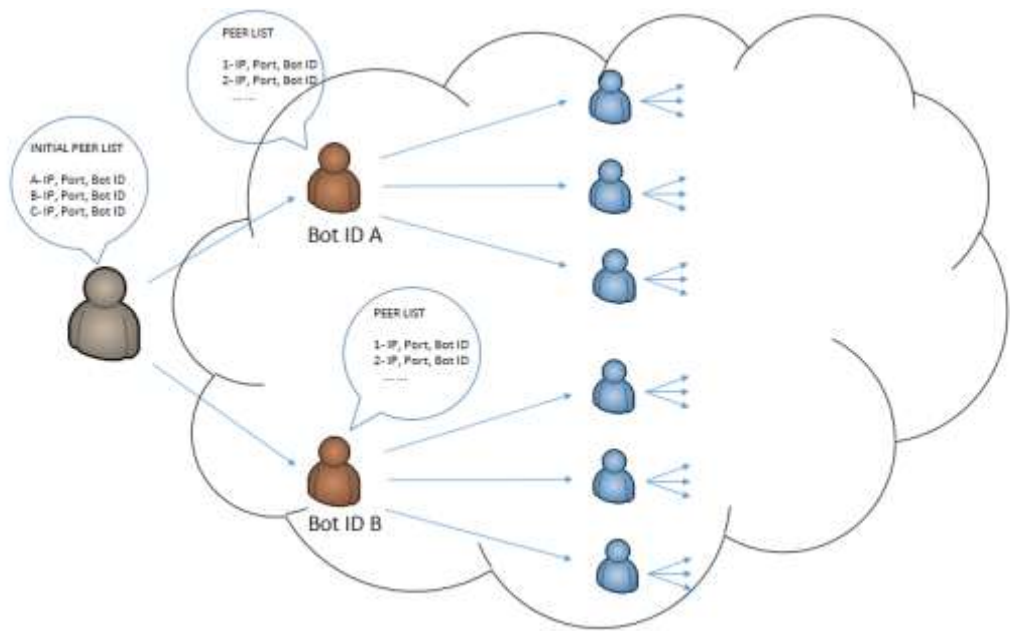

*Figure 1:a) How P2P Bot spreads*

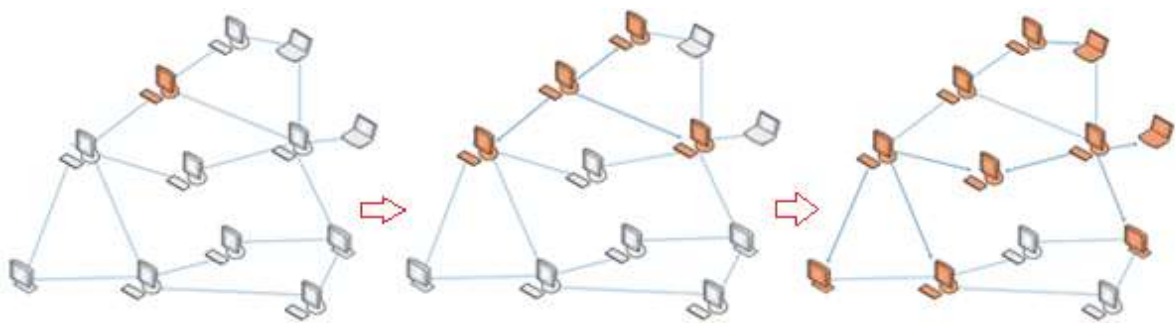

*Figure 1: b) Architecture of a P2P Botnet*

The rest of this paper is organized as follows. The previous work done on P2P botnet detection is described in the next section (section 2). Section 3 describes in detail the methodology of the author's experiment and gives a description of the algorithm used to identify the botnet activity, and section 4 describes the experimental setup that was used to conduct the research work. The paper concludes with the results and future scope of this work in section 5.

## 2. RELATED WORK ON P2P BOTNET DETECTION

Authors in *(Yen and Reiter 2010)* differentiate the network flow records based on certain features related to traffic volume and categorize them as P2P malicious and benign. They also present how the plotters could change their behavior to evade their detection technique, which was seen in observing the behavior of Nugache, which is known to randomly change its behavior.

Researchers *(Schoof and Koning 2007)* report on their work on detecting Sinit and Nugache that these bots communicate on the same port. Also a large number of destination unreachable error messages and connection reset error messages can be observed. But bots which encrypt their payloads will disable this technique from working.

Researchers *(Dittrich and Dietrich 2008)* explain the challenges and features when dealing with Nugache P2P botnet. Authors in *(Stover, et al. 2007)*conclude that it is impossible for a static Intrusion Detection System (IDS) to detect Nugache traffic. The authors' analysis of Storm concludes that the Storm bot can be detected by configuring an IDS to find the configuration file used by the bot. But it is difficult to distinguish between legitimate P2P communication and Storm bot.

Authors *(Holz, et al. 2008)*develop ways to mitigate the Storm worm and introduce an active measurement technique to enumerate the number of infected hosts by reverse engineering of the bot's binary executable, in order to identify the function which generates the key that is used for searching other infected machines and bots.

*(Stewart February 2007)* reports an in depth analysis of Peacomm and how it spams large number of emails to many accounts holding an executable attachment.

Authors in *(Grizzard, et al. 2007)* retrieve the hashes of malware and use it for locating a zombie nodes' activities in a P2P network. They argue that if a peer searches for a hash of a malware, it must be a zombie.

Authors in *(Nagaraja, et al. 2010)* devise techniques to localize Botnet members based on the unique communication patterns arising from the structured overlay topologies used for command and control. They examine if ISPs can detect the efficient communication structures of P2P bots. Their algorithm isolates this based on the information about which pairs of nodes communicate with one another.

Liu et al. in *(Liu, et al. 2010)* propose adaptive mechanisms to detect a variety of P2P Botnets. But their solutions can be applied only in the attacking stage of the Botnet. Their study was successful on Trojan Peacomm, but the implications are not clear when dealing with other categories/types of Botnets.

Authors in *(Tarng, et al. 2011)* use P2P flow identification techniques to monitor and filter traffic flows, isolating the hosts when they connect to the Botnet. Their work uses Bayes Classifier and Neural Network classifier to detect the IP address of the infected systems.

Researchers in *(Gu, et al. 2008)* present a framework named BotMiner, which detects both centralized IRC and P2P botnets using an anomaly based detection system. The assumption in this regard is that bots are coordinated malware that exhibit similar communication patterns and behaviours. BotMiner targets a group of compromised systems belonging to a monitored network, whereas it fails to detect a simple system which might be a part of a Botnet which is not in the monitored network's zone.

Researchers *(Noh, et al. 2009)* detect Botnet activity by detecting similar flows occurring between groups of hosts in the network on a regular basis. Flows with similar behaviors are labeled into groups, and a transition model of the grouped flows is constructed using a probability matrix. The authors compute a 'likelihood ratio' and use that ratio for detection of bots.

A self-organization map algorithm was applied by authors in *(Langin, et al. 2009)* to detect P2P Botnets, in which they assume that there would be numerous failed connection attempts from exterior to interior in firewall.

Most of the previous research work has focused on detecting a specific botnet activity. Such methods were not reported to be successful in detecting bots whose traffic characteristics were not used in training the machine learning algorithm. Clearly, it is seen that using a machine learning based approach is superior in detecting malicious traffic when compared to a traditional signature based approach as the bot masters redesign the bots from time to time, and the functionality, behavior etc. of the botnet varies quite significantly with each new version of the bot. Moreover previous works have not seen much research on deploying the detection module in a real-time scenario to monitor and mitigate botnet activity on a network.

## 3. Methodology

### 3.1 Feature Extraction

Machine learning algorithms require appropriate 'features' as inputs in order to train models. For this research, samples of certain P2P bots were deployed on a test-bed (as described in the next section) and network trace files (pcaps) were obtained. These trace files were then used for feature extraction using an Open-source tool Netmate *(Netmate 2011)*. Netmate gives its output in the form of 'flows' (defined by the tuple <Source IP, Source port, Destination IP, Destination port, protocol>) and extracts more than forty features applicable to each of flow. The features extracted by it are listed below. All features are represented as integers unless otherwise stated.

Table 3: Description of Features used

| Feature | Description of the feature |
| --- | --- |
| *Srcip* | *Source ip address (string)* |
| *Srcport* | *Source port number* |
| *Dstip* | *Destination ip address (string)* |
| *Dstport* | *Destination port number* |
| *Proto* | *Protocol (ie. TCP = 6, UDP = 17)* |
| *total_fpackets* | *Total packets in forward direction* |
| *total_fvolume* | *Total bytes in forward direction* |
| *total_bpackets* | *Total packets in backward direction* |
| *total_bvolume* | *Total bytes in backward direction* |
| *min_fpktl* | *Size of smallest packet sent in forward direction (in bytes)* |
| *mean_fpktl* | *Mean size of packets sent in forward direction (in bytes)* |
| *max_fpktl* | *Size of largest packet sent in forward direction (in bytes)* |
| *std_fpktl* | *Standard deviation from mean of packets sent in forward direction (in bytes)* |
| *min_bpktl* | *Size of smallest packet sent in backward direction (in bytes)* |
| *mean_bpktl* | *Mean size of packets sent in backward direction (in bytes)* |
| *max_bpktl* | *Size of largest packet sent in backward direction (in bytes)* |
| *std_bpktl* | *Standard deviation from mean of packets sent in backward direction (in bytes)* |
| *min_fiat* | *Minimum amount of time between two packets sent in forward direction (in microseconds)* |
| *mean_fiat* | *Mean amount of time between two packets sent in forward direction (in microseconds)* |
| *max_fiat* | *Maximum amount of time between two packets sent in forward direction (in microseconds)* |
| *std_fiat* | *Standard deviation from mean amount of time between two packets sent in forward direction (in microseconds)* |
| *min_biat* | *Minimum amount of time between two packets sent in backward direction (in microseconds)* |
| *mean_biat* | *Mean amount of time between two packets sent in backward direction (in microseconds)* |
| *max_biat* | *Maximum amount of time between two packets sent in backward direction (in microseconds)* |
| *std_biat* | *Standard deviation from mean amount of time between two packets sent in backward direction (in microseconds)* |
| *Duration* | *Duration of flow (in microseconds)* |
| *min_active* | *Minimum amount of time that flow was active before going idle (in microseconds)* |
| *mean_active* | *Mean amount of time that flow was active before going idle (in microseconds)* |
| *max_active* | *Maximum amount of time that flow was active before going idle (in microseconds)* |
| *std_active* | *Standard deviation from mean amount of time that flow was active before going idle (in microseconds)* |
| *min_idle* | *Minimum time a flow was idle before becoming active (in microseconds)* |
| *mean_idle* | *Mean time a flow was idle before becoming active (in microseconds)* |
| *max_idle* | *Maximum time a flow was idle before becoming active (in microseconds)* |
| *std_idle* | *Standard devation from mean time a flow was idle before becoming active (in microseconds)* |
| *sflow_fpackets* | *Average number of packets in a sub flow in forward direction* |
| *sflow_fbytes* | *Average number of bytes in a sub flow in forward direction* |
| *sflow_bpackets* | *Average number of packets in a sub flow in backward direction* |
| *sflow_bbytes* | *Average number of packets in a sub flow in backward direction* |
| *fpsh_cnt* | *Number of times PSH flag was set in packets travelling in forward direction (0 for UDP)* |
| *bpsh_cnt* | *Number of times PSH flag was set in packets travelling in backward direction (0 for UDP)* |
| *furg_cnt* | *Number of times URG flag was set in packets travelling in forward direction (0 for UDP)* |
| *burg_cnt* | *Number of times URG flag was set in packets travelling in backward direction (0 for UDP)* |
| *total_fhlen* | *Total bytes used for headers in forward direction.* |
| *total_bhlen* | *Total bytes used for headers in backward direction.* |

For these experiments, the authors removed the first four features (Source IP, Source port, Destination IP, Destination port) as they are totally dependent on the network configuration on which the bots are deployed. Information Gain Attribute Evaluation was done using the Ranker Algorithm in order find the most influential features of the entire feature set. This method evaluates the worth of an attribute by measuring the information gain with respect to the class, where Information Gain is described by the following equation:

*Information Gain (Class, Attribute) = H(Class) - H(Class | Attribute).*

The following were the first 15 in the list, with their Information gain shown in the first column.

Table 4: Features selected using Information Gain Ranking Algorithm

| Rank | Feature | | |
|---|---|---|---|
| 0.814 | total_bhlen | 0.7182 | max_bpktl |
| 0.81 | std_bpktl | 0.7104 | sflow_fbytes |
| 0.7886 | fpsh_cnt | 0.6802 | sflow_fpackets |
| 0.7751 | total_fhlen | 0.6761 | mean_fiat |
| 0.7654 | bpsh_cnt | 0.6625 | total_bpackets |
| 0.7568 | min_biat | 0.6502 | max_fiat |
| 0.7438 | min_fiat | 0.6459 | mean_biat |
| | | 0.6427 | max_fpktl |

### 3.2 Bayesian Regularized Neural Network

Artificial Neural network is a very useful tool for Machine Learning which has been applied in several scenarios. It has been applied to many cases such as text pronunciation example trained by a back propagation neural network *(Franco, et al. 1997)*. It has also found many applications in the field of pattern recognition *(Carpenter and Grossberg 1988)*. In predictive modeling, application of the Artificial Neural Networks has the advantage of being able to capture relationships which are highly complex.

Neural network architecture, A, consists of a specification of the number of layers, the number of units in each layer, the type of activation function performed by each unit, and the available connections between the units. Values for the weight, w, is assigned to the connections in the network, the weighted input, x, sum is mapped with y1(x; w, A), the predicted value of the output. The distance of the predicted value to the training set is measured by some error function. The error for the entire data set is commonly taken to be

$$E(D,w,A) = \sum_{m=0}^{N} \frac{[y1(x^m;w;A)-y^m]^2}{2} \quad (1)$$

Here E is the error function, often called as Mean of Squares of errors (MSE), y1 is the predicted output, x is the input and m is the instance of the data. Here y represents the output data and N represent the total data set.

It has been observed in literature that back propagation neural networks used to train the neural network model give high reliability *(Zhang, et al. 2001)*. The back propagation learning algorithm uses a gradient search technique to minimize the mean square error of the output of the network.

The parameters of the back propagation networks are generally set by trial and error. Reserved test data is used to assess its generalization ability (or the cross-validation ability) *(MacKay 1992)*. These parameters change the effective learning model, for example, number of hidden units, weight decay terms etc.

The training of the network is to find a set of weights, w, that gives optimal map between the training set and the predicted set. The learned weights are expected to fit well to new examples. Plain back-propagation learns by performing gradient descent to optimize the error function. More efficient optimization techniques may also be used, such as conjugate gradients or variable metric methods. Figure 2 shows the method of selecting optimal number of hidden neuron using the correlation coefficients at each stage.

However there are certain disadvantages of using back propagation networks:

    1. It can over fit the given data.

    2. It needs large datasets to correctly map the relationship between the input and outputs.

As explained above, it is seen that although back propagation neural networks are very reliable, they often suffer with the problem of over fitting the experimental data. Over-fitting problem or poor generalization of the given dataset happens when a network over-learns during the training period. The result is a "too well-trained" model, which may not perform well on unseen data. This problem is known as Occam's razor problem *(MacKay 1992)*. The principle is that unnecessarily complex models should not be preferred than the simpler ones. In order to address this issue, the method of 'Bayesian inference' is used which automatically addresses the Occam's razor problem.

According to the fundamentals of the Bayesian analysis, the plausibility of alternative hypothesis is represented by probabilities, and inference is performed by evaluating those probabilities. Thus, using Bayesian probability theory it is possible to automatically infer the flexibility of a model warranted by the data. The model can be evaluated by using the simple Bayes rule given by

$$P\left(H_i/D\right) = \frac{P(H_i)P(D/H_i)}{P(D)} \quad (2)$$

Where the denominator P (D) is the normalizing constant which makes the final belief add up to 1.

Bayesian Regularization approach minimizes the over-fitting problem by taking into account the 'goodness-of-fit' as well as the network architecture. Hence in this work the Bayesian regularization approach is used for better fitness of the data. The neural network toolbox of MATLAB software package is used for training and testing the given data. Levenberg-Marquardt algorithm with Bayesian regularization function (trainbr) has been used to train the network. The network architecture is obtained for which the network has minimum sum of squares of errors (SSE) and also has better generalization ability.

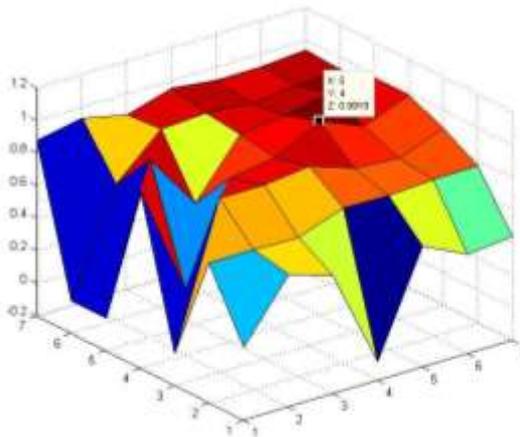

*Figure 2: Selecting optimal number of hidden neurons for BR-ANN*

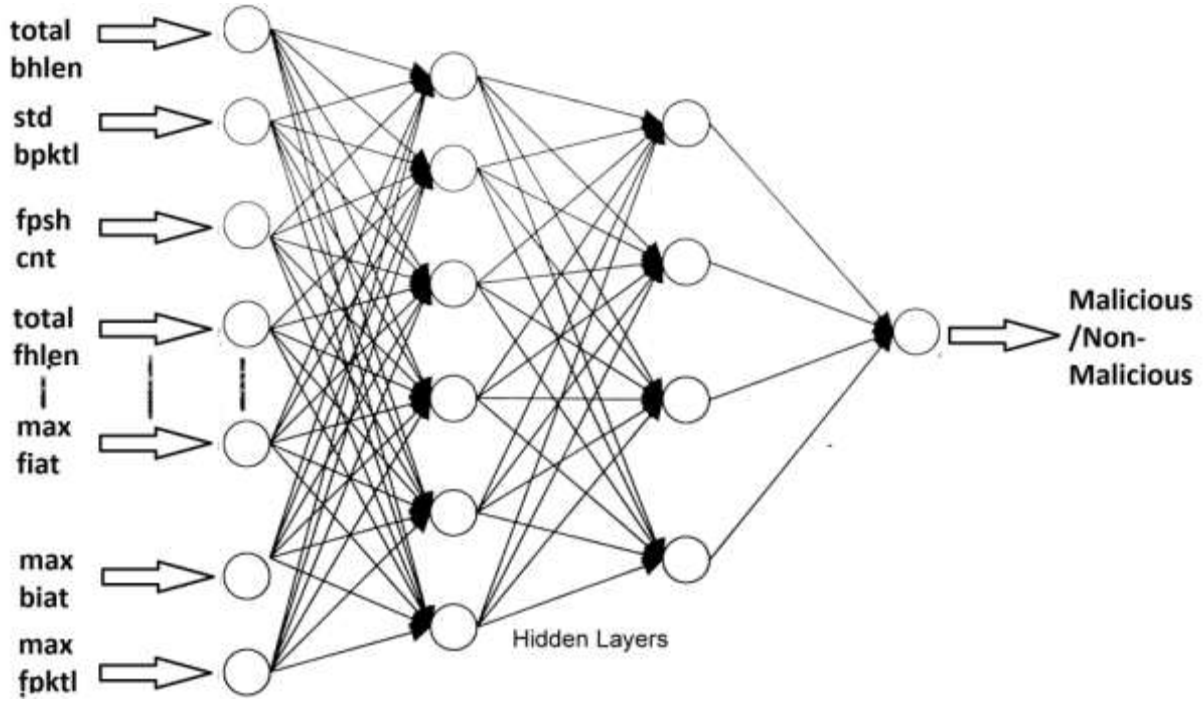

*Figure 3: Architecture of Neural Network for Botnet Detection*

The fifteen features are the input, and output is the class label indicating the nature of the flow. The input is first normalized from 0.05 to 0.95 using the following equation for transfer function to activate efficiently as the tan sigmoid function is used for mapping from the weighted inputs to the output.

$$x_n = 0.05 + 0.9 * (x - x_{min})/(x_{max} - x_{min})$$

Where $x_{min}$ and $x_{max}$ are the minimum and maximum values of x and $x_n$ is the normalized data of the corresponding x. Once the best trained network is found, all the transformed data returns to their original value using the following equation:

$$x = x_{min} + (x_n - 0.05) * (x_{max} - x_{min})/0.9$$

### 3.3 Algorithm used and pseudo-code

The entire code is divided into two modules. One module is the getConversations method and the other is testtshark method. getConversations reads from the specified pcap files and invokes Netmate to extract flow statistics from the pcap file which are used to build the BR-ANN model. Testtshark is the real-time module which uses jnetpcap to extract the same features out of live traffic and submits the flows to the Matlab-Weka-API for classification.

**getConversations()**
___________________________________________________________________
1. BEGIN
2. invoke Netmate
    a. packet properties are extracted and buffered
    b. packets are characterised as flows and their 44 statistics are
       calculated
    c. statistics are written into csv file
3. csv file converted to arff file by adding Attribute and Data headers
4. Instances in the arff file divided into 90% training data and 10%
   testing data; sent to the MATLAB BR-ANN code as input to train the
   model
5. Trained model is imported into Weka using MATLAB-WEKA API
6. END

**testtshark()**
___________________________________________________________________
1. BEGIN
2. packet capture started on the wire using tshark
3. jnetPcap invoked
    a. get packet properties of every packet
    b. send the packets to a buffer
    c. group on <source_ip, destination_ip, source_port,
       destination_port, protocol> to form flows.
    d. Calculate flow statistics for each of the keys in the above
       step
    e. Write the flow statistics to another buffer converting them
       into instances of test data
4. Send the Instances to Weka API to test against the loaded training
   model
    a. If instance is malicious
         i. Flag the class label to be malicious
        ii. Make a note of the time stamp
       iii. Search for the packet containing this time stamp and
            write it to a separate pcap file.
    b. If instance is non-malicious
         i. Ignore and continue
5. Send the pcaps classified as malicious to an IDS/IPS/Firewall for
   further actions
6. END

## 4. Experimental Setup

### 4.1 Collecting Dataset

The test-bed for this research work consists of a standalone network of Linux systems. The systems were connected to an Access switch in order to form the standalone network. On top of each of these physical machines, the authors ran virtual machines with Windows XP as the operating system. On these virtual machines, samples of Kelihos-Hlux, Zeus, Waledac were deployed. This setup is shown in figure 4. The samples of these malware were obtained from (contagiodump.blogspot.in and openmalware.com). The network activity of these malware samples were monitored for 48 hours each and the activity was captured by Wireshark. After collecting packet traces of the network activity of each of the malware, right from the infection stage to the attack stage, the pcap files were stored in the database server for further analysis.

The 'malicious dataset' contained a total of 55,824 flows. In order to achieve proper classification, benign traffic consisting of traffic from p2p applications, ftp transfers, telnet sessions, video streaming and mobile updates was collected by the authors, and flows were extracted in the same manner as for malicious traffic. The entire dataset was merged and given to the machine learning algorithm for model generation (described in Section 3.2).

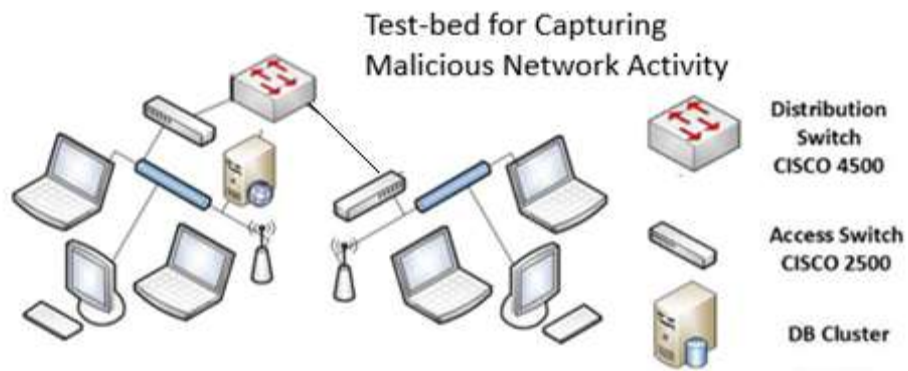

*Figure 4: Testbed for Capturing Malicious Network Activity*

## *4.2 Real-time Botnet Detection*

For real-time detection of botnet activity in the network, a module was deployed on a system which was receiving mirrored traffic from the entire network, as shown in figure 5. This module utilizes tshark to read the packets and stores them in libpcap format in chunks of 200 MB each. The stored packet captures are converted into conversations (or flows). Each flow is an instance which is to be monitored for malicious activity.

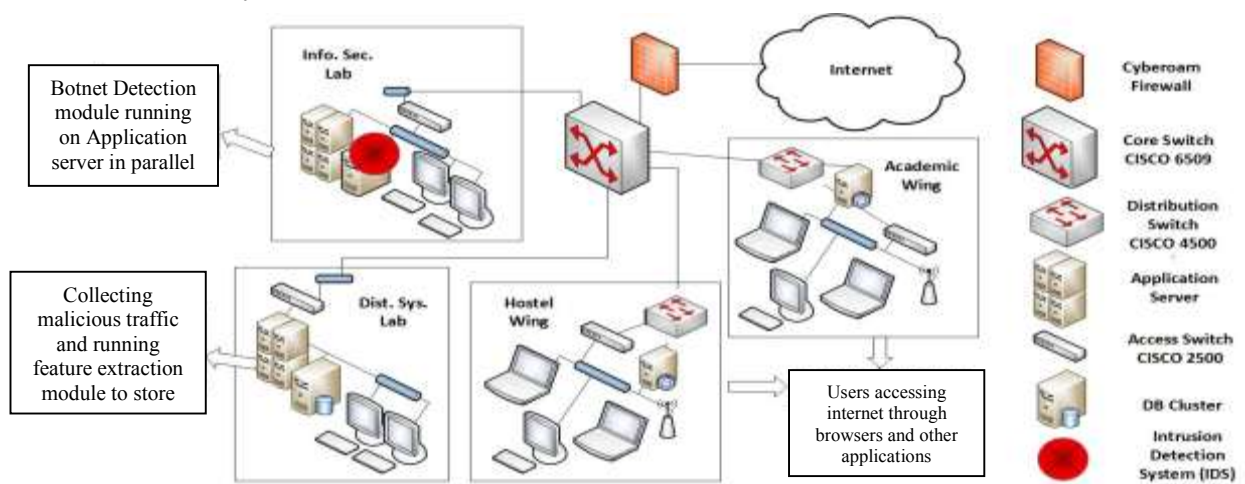

*Figure 5: The test-bed used for deploying the P2P Botnet Detection Module*

The Bayesian Neural Network model, whose training was described in the previous section on methodology, was inherited using Matlab-Weka Interface API in order to test the flows against the trained model in real-time. According to the classification done by the trained model, only those flows (and thus only those packets) which are marked malicious by the algorithm are saved into a separate pcap file. All this is done in real-time thereby giving an indication on the botnet activity in a network's present state. This module, shown in figure 6, can serve as a good pre-processing module for any IDS/IPS to detect Botnet traffic and flag an alert for the same. The working snapshot of the model is shown in figure 7, running on eclipse IDE.

| LibPcap Network Capture | | |
|---|---|---|
| P2P Bot | Flow Extraction | P2P and other Benign |

| Building Br-ANN Model | |
|---|---|
| Training +Testing Sets | Bayesian ANN Model |

| Real-Time Detection |
|---|
| Java Module for P2P Bot Detection |

*Figure 6: Overview of Building the BR-ANN Module*

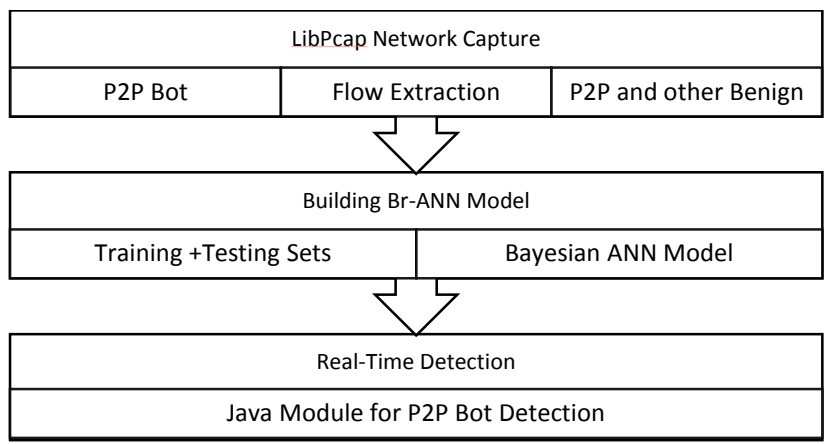

*Figure 7: Screenshot of Real-Time P2P Botnet Detection Module in Action*

## 5. Results and Future Scope of Work

The exact sequence of steps used in this real-time P2P Botnet detection module are shown in figure 8:

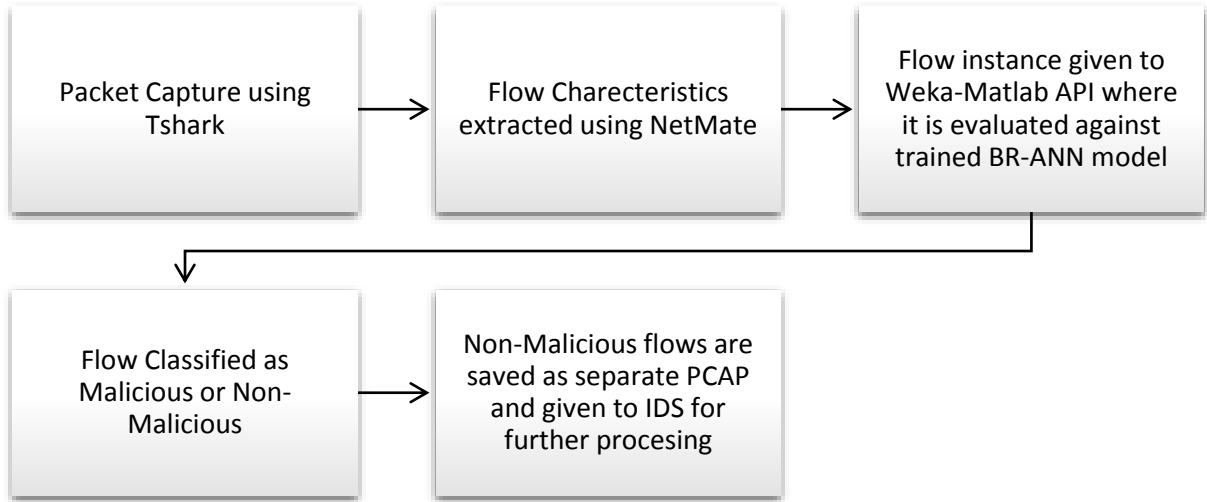

Figure 8: Overview of Real-time Deployment of the Detection Module

Correlation between two variables X and Y is measured using the Pearson product-moment coefficient, which takes the value between -1 and +1 inclusive. It is defined by the formula:

$$r = \frac{\sum_{i=1}^{n}(X_i - \bar{X})(Y_i - \bar{\bar{Y}})}{\sqrt{\sum_{i=1}^{n}(X_i - \bar{X})^2}\sqrt{\sum_{i=1}^{n}(Y_i - \bar{Y})^2}}$$

The ideal prediction is supposed to generate a straight line passing through origin at an angle of 45 degrees, as the X-axis and Y-axis represent the experimental and predicted values by each of the methods employed. It is observed that the correlation coefficient is very near to 1. This shows that the goodness of fit of the model generated by Bayesian Regularized ANN is close to exact prediction.

90% of the data is considered as training data and remaining 10% is considered testing data to validate the goodness of fit of the model generated. The correlation obtained by the network trained is very high and is equal to 0.9931. When tested on bots which were not used in the training set, with samples obtained from CAIDA (The CAIDA UCSD Network Telescope "Three Days Of Conficker" -21st December 2009) and ISOT (Saad, et al. July 19-21, 2011), the BR-ANN model gave a correlation of 0.9902. These results are given below in the box plot and the histogram respectively in figures 9 and 10. A box plot can be used to graphically show groups of quantitative data through their five number summaries: the smallest observation (sample minimum), lower quartile (Q1), median (Q2), upper quartile (Q3), and largest observation (sample maximum). C1 depicts the deviation in experimental results and C2 depicts the deviation in the predicted results. From the histogram it can be seen that the predicted number of instances in non-malicious and malicious is very close to the experimental number of instances and thus these two establish the working of the BR-ANN module. The scatter plot depicts the precision-recall characteristics of the model where Recall is on the x-axis and Precision on the y-axis. Recall is the fraction of positive examples that are correctly labeled, whereas Precision measures that fraction of examples classified as positive that are truly positive. Statistical tests of F-test and Levene's test were performed in order to judge about the variance of the predicted values of the model compared with the experimental data. The results of the tests performed are provided in Tables. The significance value, p is observed to be more than 0.05 for 95 per cent confidence interval. Consequently it can be said that the Null hypotheses cannot be rejected. This means that the predictions of the Neural Network have satisfactory mapping to the experimental data. All the p-value are almost 0.9, indicating the model has good accuracy with the capability of generalization. This shows that the goodness of fit of the model generated by Bayesian Regularized ANN is statistically satisfactory and accurate.

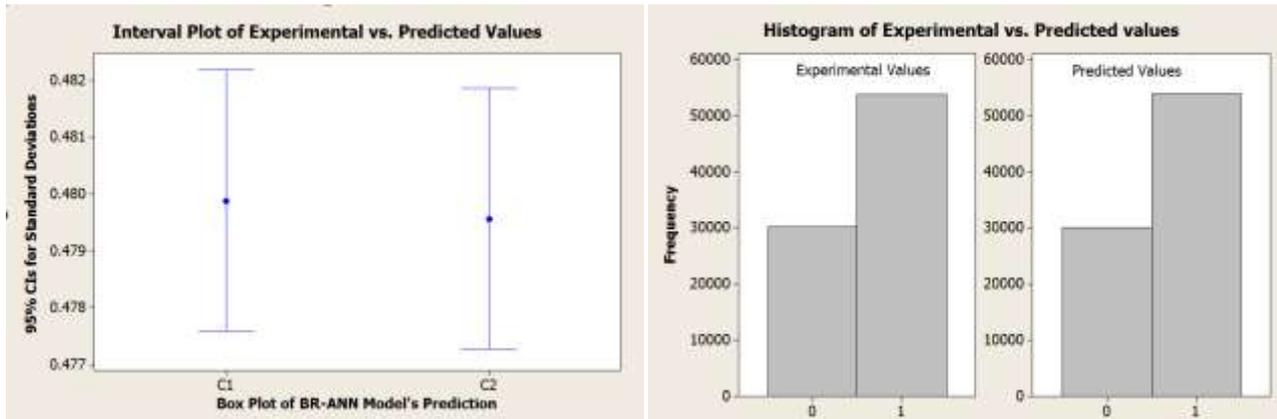

Figure 9: Accuracy measure of BR-ANN Model a) Box Plot b) Histograms

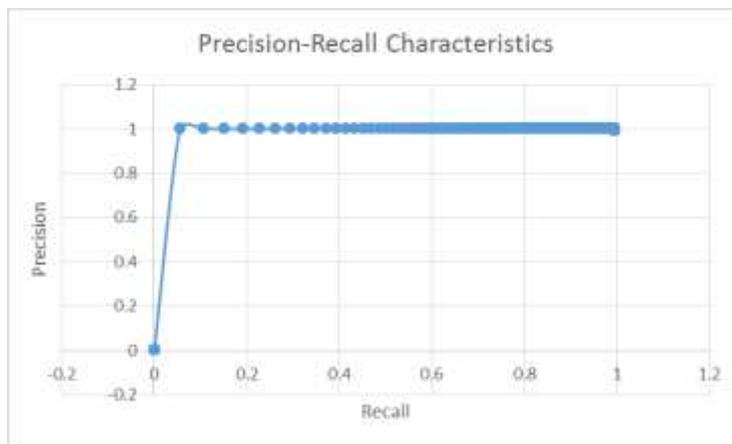

Figure 10: Precision-Recall Plot of the Testing Data

Table 5: Statistical Test Values

| Method | DF1 | DF2 | Statistic | P-Value |
|---|---|---|---|---|
| F Test (normal) | 75628 | 75628 | 1 | 0.847 |
| Levene's Test (any continuous) | 1 | 151256 | 0.22 | 0.641 |

The prediction results were also verified with the logs of Cyberoam, a Unified Threat management system deployed at the authors' organization and which is also in use in many corporate organizations. The packets flagged as malicious by the Bayesian Neural Network Detection Module were also flagged by Cyberoam to be malicious, as shown in figure 11. This shows that the module build on BR-ANN can be built as an efficient pre-processing engine for existing IDS/IPS solutions.

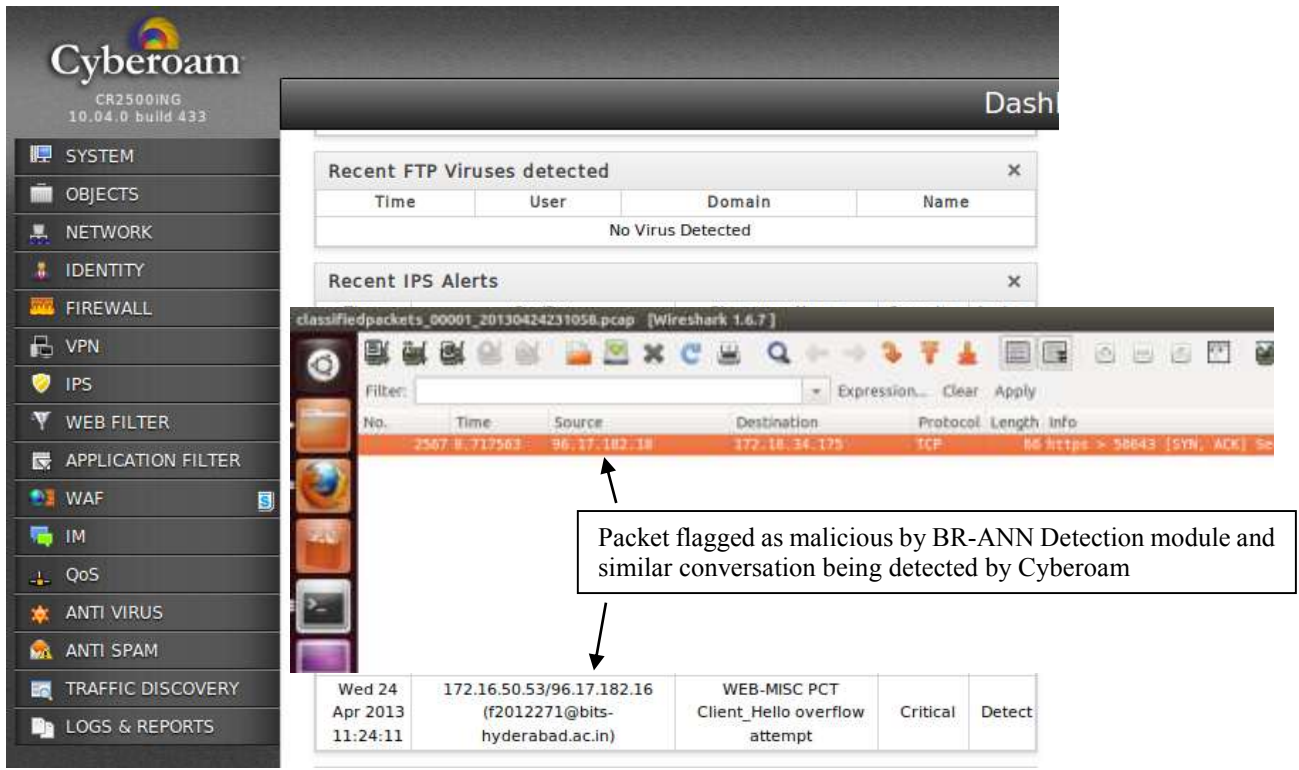

*Figure 11: BR-ANN detection module's results verified by the UTM Cyberoam*

In this research, existing models of Machine Learning, in the context of P2P botnet detection, were studied. And a better model for Artificial Neural Networks based on Bayesian Regularization was proposed, which is very efficient in the problem context which need good generalization abilities. Thus it is conclusively shown through the statistical tests that the trained BR-ANN model is able to generalize very well and is able to predict the activity of unknown bots' malicious activity.

Detecting specifically the kind of malicious activity and thereby identifying the unknown bot could help the security experts to take appropriate preventive measures. Due to the large bandwidth of the network, shifting to scalable and distributed architecture is a very attractive alternative to consider. Integrating the existing API with Mahout and Hadoop APIs and re-writing the classifier to suit the Map-Reduce paradigm is the future scope of this work.

# REFERENCES

Apvrille, A. 2012. "Symbian worm Yxes: Towards mobile botnets?" *Journal in Computer Virology*, 117-131.

2012. *BBC News*. December 10. http://www.bbc.co.uk/news/mobile/technology-11968605.

Carpenter, G. A., and S. Grossberg. 1988. "The ART of adaptive pattern recognition by a self-organizing neural network." *Computer* 77-88.

Dittrich, D., and S. Dietrich. 2008. "P2P as botnet command and control: a deeper insight." *International Conference on Malware*. IEEE. 41-48.

Franco, H., L. Neumeyer, Y. Kim, and O. Ronen. 1997. "Automatic pronunciation scoring for language instruction." *IEEE International Conference on Acoustics, Speech, and Signal Processing*. IEEE. 1471-1474.

Grizzard, J. B., V. Sharma, C. Nunnery, B. B. Kang, and D Dagon. 2007. "Peer-to-peer botnets: Overview and case study." *First Workshop on Hot Topics in Understanding Botnets*. 1-1.

Gu, G., R. Perdisci, J. Zhang, and W. Lee. 2008. "BotMiner: Clustering analysis of network traffic for protocol-and structure-independent botnet detection." *17th conference on Security symposium*. 139-154.

Holz, T., M. Steiner, F. Dahl, E. Biersack, and F Freiling. 2008. "Measurements and mitigation of peer-to-peer-based botnets: a case study on storm worm." *1st Usenix Workshop on Large-Scale Exploits and Emergent Threats*. 1-9.

Langin, C., H. Zhou, S. Rahimi, B. Gupta, M. Zargham, and M. R. Sayeh. 2009. "A self-organizing map and its modeling for discovering malignant network traffic." *IEEE Symposium on Computational Intelligence in Cyber Security*. IEEE.

Liu, D., Y. Li, Y. Hu, and Z. Liang. 2010. "A P2P-botnet detection model and algorithms based on network streams analysis." *International Conference on Future Information Technology and Management Engineering (FITME)*. 55-58.

MacKay, D. J. 1992. "Bayesian interpolation. ." In *Neural computation*, by D. J. MacKay.

Nagaraja, S., P. Mittal, C. Y. Hong, M. Caesar, and N. Borisov. 2010. "BotGrep: finding P2P bots with structured graph analysis." *19th USENIX conference on Security*. USENIX Association. 7-7.

2011. *Netmate*. August.

Noh, S. K., J. H. Oh, J. S. Lee, B. N. Noh, and H. C. Jeong. 2009. "Detecting P2P botnets using a multi-phased flow model." *Third International Conference on Digital Society*. IEEE. 247-253.

Saad, Sherif, Issa Traore, Ali A. Ghorbani, Bassam Sayed, David Zhao, Wei Lu, John Felix, and Payman Hakimian. July 19-21, 2011. "Detecting P2P botnets through network behavior analysis and machine learning." *9th Annual Conference on Privacy, Security and Trust*. Montreal, Quebec, Canada.

Schoof, R., and R. Koning. 2007. *Detecting peer-to-peer botnets*. University of Amsterdam.

Stewart, J. February 2007. "Peacomm DDOS Attack."

Stover, S., D. Dittrich, J. Hernandez, and S. Dietrich. 2007. "Analysis of the Storm and Nugache trojans: P2P is here." *USENIX*. 18-27.

Tarng, W., L. Z. Den, K. L. Ou, and M. Chen. 2011. "The Analysis and Identification of P2P Botnet's Traffic Flows." *International Journal of Communication Networks and Information Security (IJCNIS)*.

2009. *The CAIDA UCSD Network Telescope "Three Days Of Conficker" -21st December*. http://www.caida.org/data/passive/telescope-3days-conficker_dataset.xml.

Wang, P., L. Wu, B. Aslam, and C. C. Zou. 2009. "A systematic study on peer-to-peer botnets." *Computer Communications and Networks*. 1-8.

Wood, D. 2010. "The skype is no longer the limit–new ways malware keeps in touch with your friends."

Yen, TF., and M.K. Reiter. 2010. "Are your hosts trading or plotting? telling p2p file-sharing and bots apart." *Distributed Computing Systems*. IEEE. 241-252.

Zhang, Z., J Li, C. N. Manikopoulos, J. Jorgenson, and J. Ucles. 2001. "HIDE: a hierarchical network intrusion detection system using statistical preprocessing and neural network classification." *IEEE Workshop on Information Assurance and Security*. 85-90.